\documentclass[12pt]{article}

\makeatletter
 
\@addtoreset{equation}{section}
\makeatother

\newcommand{\hs}{\hspace}
\newcommand{\ts}{\hspace*}
\newcommand{\vs}{\vspace}
\newcommand{\e}{\enskip}
\newcommand{\q}{\quad}

\newcommand{\dps}{\displaystyle}
\newcommand{\f}{\frac}

\newcommand{\Phat}{\hat{{\cal P}}}
\newcommand{\Qhat}{\hat{\Omega}}

\newcommand{\Zp}{\hat{Z}^{(+)}}
\newcommand{\Zm}{\hat{Z}^{(-)}}
\newcommand{\Zpm}{\hat{Z}^{(\pm)}}
\newcommand{\Zmp}{\hat{Z}^{(\mp)}}

\newcommand{\xipps}[2]{\hat{\xi}^{\mbtn{#1}(+)}_{#2}}

\newcommand{\pipps}[2]{\hat{\pi}^{\mbtn{#1}(+)}_{#2}}

\newcommand{\hope}{{\it hyper}-operator}
\newcommand{\bM}{\bar{M}}
\newcommand{\mcK}{\mathcal{K}}
\newcommand{\mcC}{\mathcal{C}}
\newcommand{\mcA}{\mathcal{A}}

\newcommand{\mcS}{\mathcal{S}}
\newcommand{\mcH}{\mathcal{H}}
\newcommand{\mcO}{\mathcal{O}}

\newcommand{\commu}[2]{[#1,#2]}
\newcommand{\commut}[2]{[#1,\e #2]}
\newcommand{\symmp}[2]{\ \{#1,#2\}}
 
\newcommand{\mbtn}[1]{\mbox{{\tiny #1}}}
\newcommand{\pscrp}{\mbox{{\scriptsize $\Phat$}}}

\begin{document}
\begin{center}
{\bf \Large Canonical Structure of Noncommutative Quantum Mechanics as Constraint System}\vs{12pt}\\
M. Nakamura\footnote{Present address:Research Institute, Hamamatsu Campus, Tokoha University,\\
\ts{18pt}E-mail:mnakamur@hm.tokoha-u.ac.jp
}\vs{12pt}\\
{\it Department of Service and Business, Hamamatsu University, Miyakoda-cho 1230, 
Kita-ku, Hamamastu-shi, Shizuoka 431-2102, Japan}
\end{center}
\vs{24pt}

\begin{abstract}
Starting with the first-order singular Lagrangian, the canonical structure in the noncommutative quantum mechanics with the noncommutativities both of coordinates and momenta is investgated. Using the projection operator method (POM) for the constraint systems and the constraint star-product, the noncommutative quantum system is constructed and the commutator algebra of {\it projected} canonically conjugate set(CCS) of the system is derived in the form including all orders of the noncommutativity parameters. We discuss the alternative CCS, which obeys the ordinary noncommutative commutator algebra. The {\it exact} CCS is constructed in the framework of the POM, and which is shown to be equivalent to the CCS constructed through the Seiberg-Witten map and the Bopp shift. We further discess the alternative Lagrangian to realize the noncommutativities both of coordinates and momenta. 
\end{abstract}

%\tableofcontents

\section{Introduction}	

\ts{12pt}Since Snyder\cite{Sny}, Noncommutative extensions of quantum mechanics and quantum field theories have been extensively investigated until now\cite{SW,Conne,Szabo,Kokado,Li,Bastos,Kupriyanov}. It is well known that the dynamical systems with the noncommutativity among  coordinates, momenta are able to be descrived by the constraint systems, and the dynamical models for such constraint systems have been investigated widely\cite{GJ, Jackiw, Drglzv, Grti, BmfGrot, DrgRak, Plyush1, Plyush2, Plyush3, Plyush4, stchl, Plyush5, Plyush6, Drglzv2, Bnrj}, and it has been shown that Chern-Simons like terms play the important role in the realization of noncommutativities\cite{ncqm}.\\
\ts{12pt}Using the projection operator method (POM) with the star-product quantization\cite{MN1,MN2,Kr1,Kr2}, in this paper, we shall construct the noncommutative quantum system with the noncommutativities both of coordinates and momenta in more general form.\\
\ts{12pt}For this purpose,  we propose the first-order singular model  Lagrangian with two kinds of Chern-Simons like terms, and the initial unconstraint quantum system containing the sets of second-class constraint operators is prepared. The final  
 constraint quantun system will be constructed through the successive projection-procedure\cite{MN4}. Then, the commutator algebra of the {\it projected} canonically conjugate set (CCS) constisting of coorninate operators $q^i$ and momentum ones $p_i$ $(i=1,\cdots,N)$ will be shown to take the following form:
$$
\commu{q^i}{q^j}=i\hbar(M^{-1}\Theta M^{-1})_{ij},\hs{6pt}\commu{q^i}{p_j}=i\hbar(M^{-1}\bM M^{-1})_{ij},\hs{6pt}\commu{p_i}{p_j}=i\hbar(M^{-1}\Xi M^{-1})_{ij},
\eqno{(1.1)}
$$
where $M$ and $\bM$ are the $N\times N$ matrices, 
$$
M=I+\f14\Theta\Xi,\hs{12pt}\bM=I-\f14\Theta\Xi
\eqno{(1.2)}
$$
with the $N\times N$ unit matrix $I$, and $\Theta^{ij}$, $\Xi_{ij}$, the $N\times N$ totally antisymmetric  matrices. These matrices will be defined in Sect.3 and their algebraic properties, discussed there. \\
As shown in Ref.\cite{MN4}, the results of the successive projections generally depend on the order of the operations of projection operators. Because of the structure of the constraint operators in the initial system, however, the commutator algebra of the final ptojected system will be shown to be independent to the order of projections. \\
\ts{12pt}Following the POM, we will construct the CCS consisting of $Q^i$ and $P_i$ $(i=1,\cdots,N)$ in terms of the {\it projected} CCS, which satisfies
$$
\commu{Q^i}{Q^j}=0,\hs{6pt}\commu{Q^i}{P_j}=i\hbar\delta^i_j,\hs{6pt}\commu{P_i}{P_j}=0,
\eqno{(1.3)}
$$
and which we shall call the {\it exact} CCS.
Then, it will be shown that the {\it exact} CCS provided by the POM holds the equivalent structure to the CCS obtained by the Seiberg-Witten map\cite{SW} and the Bopp shift\cite{Bopp1}. We further discuss the alternative model Lagrangian to realize the noncommutativities both of coordinates and momenta.  
 \\
\ts{12pt}This paper is organized as follows. In Sect.2, we briefly review the POM with the star-product quantization. In Sect.3, we first discuss  the algebraic properties of the antisymmetric mateices $\Theta,\Xi$. We next propose the model Lagrangian and the initial unconstraint quantum system. Then, the canonical structure of the final constraint quantum system is constructed, and the alternative {\it projected} CCS is proposed. Following the POM, in Sect.4, we construct the {\it exact} CCS satisfying the commutator algebra (1.3) and the unified expression for the projected Hamiltonian of the final constraint quantum system is given. In Sect.5, we mention the alternative model Lagrangian, and some concluding remarks are given. Furthermore, we propose the different type of constraint dynamical model, which does not contain the redundant CCS, in Appendix E.

\section{Star-product Quantization}

\ts{12pt}We here present the brief review of the POM of constraint systems including the supersymmetry with the star-product quantization\cite{MN1,MN2}. \\ 
\ts{12pt}Let $\mcS=(\mcC,\mcA(\mcC),H(\mcC),\mcK)$ be the initial unconstraint quantum system with the graded commutator algebra\footnote{For any operators $A,B$, the graded commutator, $\commu{A}{B}=AB-(-1)^{\epsilon(A)\epsilon(B)}BA$, 
 and the graded symmetrized product, $\symmp{A}{B}=\f12(AB+(-1)^{\epsilon(A)\epsilon(B)}BA)$. }, where $\mcC=\{(q^i,p_i);i=1,\cdots ,N\}=\mcC(q,p)$ is a set of canonically conjugate operators (CCS),\footnote{We shall denote $\mcC$ with $\mcC(q,p)$ and any $\mcO$ with $\mcO(q,p)$ when needed.}$\mcA(\mcC)$, the commutator algebra of $\mcC$ with  
$$
\mcA(\mcC):\commu{q^i}{p_j}=i\hbar\delta^i_j,\hs{36pt}\commu{q^i}{q^j}=\commu{p_i}{p_j}=0,
\eqno{(2.1)}
$$ 
and $H(\mcC)=H(q,p)$ is the Hamiltonian of the initial unconstraint system, $\mcK=\{T_{\alpha}(\mcC)|\alpha=1,\cdots ,2M<2N\}$, the set of the constraint-operators $T_{\alpha}(\mcC)$ corresponding to the second-class constraints $T_{\alpha}\approx 0$. Starting with $\mcS$, our goal is to construct the constraint quantum system  $\mcS^*=(\mcC^*,\mcA^*(\mcC^*),H^*(\mcC^*))$, where  $\mcC^*$, which we shall call the {\it projected} CCS, is the set of $N-M$ projected canonically conjugate pairs stisfying
$$
T_{\alpha}(\mcC^*)=0\hs{48pt}(\alpha=1,\cdots,2M).
\eqno{(2.2)}
$$
\ts{12pt}The first step is to construct the associated canonically conjugate set (ACCS) from the constraint-operators $T_{\alpha}(\mcC)$ and to prepare the projection operator $\Phat$ to eliminate $T_{\alpha}$ in the system, that is, $\Phat T_{\alpha}=0\ (\alpha=1,\cdots,2M)$, which we shall call the projection conditions\cite{MN1}.\\
\ts{12pt}Due to the Darboux's theorem in the dynamical systems, it is possible in general to define the ACCS. 
 Let $\{(\xi^a,\pi_ a)|\epsilon(\xi^a)=\epsilon(\pi_a)=s,a=1,\cdots,M\}$ be the ACCS, and their symplectic forms be
$$ 
Z_{\alpha}=\left\{\begin{array}{l}\xi^a\hs{24pt}(\alpha=a)\vs{6pt}\\
\pi_a\hs{24pt}(\alpha=a+M) \hs{36pt}(\alpha=1,\cdots,2M\q;\q a=1,\cdots,M),
\end{array}\right.
\eqno{(2.3)}
$$
which obey the commutation relation
$$
\commu{Z_{\alpha}}{\ Z_{\beta}}=i\hbar(- (-)^s)J_{\alpha\beta}=i\hbar J^{\alpha\beta},
\eqno{(2.4)}
$$
where $s=\epsilon(\xi^a)=\epsilon(\pi_a)$ is the Grassmann parity of $\xi,\pi$ and
$J^{\alpha\beta}$ is the symplectic matrix and $J_{\alpha\beta}$ is the inverse of $J^{\alpha\beta}$. Then, we define the symplectic \hope s $\Zpm_{\alpha}(\alpha=1,\dots,2M)$ as follows:
$$
\Zm_{\alpha}=\f1{i\hbar}\commu{Z_{\alpha}}{\q},\hs{36pt}\Zp_{\alpha}=\symmp{Z_{\alpha}}{\q},
\eqno{(2.5)}
$$
which, from (2.4), obey the {\it hyper}-commutation relations
$$
\begin{array}{l}
[\Zpm_{\alpha},\Zpm_{\beta}]=0,\vs{6pt}\\
\commu{\Zpm_{\alpha}}{\Zmp_{\beta}}=\commu{\Zmp_{\alpha}}{\Zpm_{\beta}}=J^{\alpha\beta}.
\end{array}
\eqno{(2.6)}
$$
The projection operator $\Phat$ is defined by\cite{MN1}
$$
\Phat=\exp\left[(-1)^s\Zp_{\alpha}\f{\partial}{\partial\varphi_{\alpha}}\right]\exp[J^{\alpha\beta}\varphi_{\alpha}\Zm_{\beta}]|_{\phi=0},
\eqno{(2.7)}
$$  
and the projection conditions for $\Phat$ are represented by
$$
\Phat T_{\alpha}(\mcC)=T_{\alpha}(\Phat\mcC)=0\hs{48pt}(\alpha=1,\cdots,2M),
\eqno{(2.8a)}
$$
which we shall briefly denote as\footnote{For a set of operators $\mcO\{O_n|n=1,2,\cdots\}$, we hereafter represent $\Phat O_n$, $O_n(\mcC)$ $(n=1,2,\cdots)$ as $\Phat\mcO$, $\mcO(\mcC)$, respectively.}
$$
\Phat\mcK(\mcC)=\mcK(\Phat\mcC)=0.
\eqno{(2.8b)}
$$ 
\ts{12pt}We next introduce two kinds of star-product as follows\cite{MN2}:
For any operators $X$ and $Y$,
$$
X\star Y=\left.\exp(\f{\hbar}{2i}\Omega_{\eta\zeta})X(\eta)Y(\zeta)\right|_{\eta=\zeta}
\eqno{(2.9)}
$$  
and
$$
X\pscrp\star   Y=\left.\left(\Phat(\eta)\Phat(\zeta)\exp(\f{\hbar}{2i}\Qhat^t_{\eta\zeta})X(\eta)Y(\zeta)\right)\right|_{\eta=\zeta}.
\eqno{(2.10)}
$$
Here, $\Qhat_{\eta\zeta}$ is the graded {\it hyper}-operator defined by
$$
\Qhat_{\eta\zeta}=(-1)^sJ^{\alpha\beta}\Zm_{\alpha}(\eta)\Zm_{\beta}(\zeta)
\eqno{(2.11a)}
$$
with the nonlocal representations for the operations of {\it hyper}-operators\cite{MN2}, and
$$
\Qhat^t_{\eta\zeta}=\Qhat_{\zeta\eta}=-\Qhat_{\eta\zeta}.
\eqno{(2.11b)}
$$
\ts{12pt}Using the $\star$ and $\pscrp\star$-products, we finally define the commutator-formulas and the symmetrized product-ones under the operation of $\Phat$ as follows:
$$
\begin{array}{lcl}
\commu{\Phat X}{\Phat Y}&=&\Phat\commu{X}{Y}_{\star}=\Phat(X\star Y-(-1)^{\varepsilon_X\varepsilon_Y}Y\star X),\vs{12pt}\\
\symmp{\Phat X}{\Phat Y}&=&\Phat\symmp{X}{Y}_{\star}=\dps{\f12}\Phat(X\star Y+(-1)^{\varepsilon_X\varepsilon_Y}Y\star  X),
\end{array}
\eqno{(2.12)}
$$  
and
$$
\begin{array}{lcl}
\Phat\commu{X}{Y}&=&\commu{X}{Y}_{\pscrp\star}=(X\pscrp\star Y-(-1)^{\varepsilon_X\varepsilon_Y}Y\pscrp\star X),\vs{12pt}\\
\Phat\symmp{X}{Y}&=&\symmp{X}{Y}_{\pscrp\star}=\dps{\f12}(X\pscrp\star Y+(-1)^{\varepsilon_X\varepsilon_Y}Y\pscrp\star X).
\end{array}
\eqno{(2.13)}
$$

\section{Construction of Noncommutative Quantum System} 

\ts{12pt}We shall consider the dynamical model to  
realize both of space-space and momentum-momentum noncommutativities with the constant noncommutativity-parameters. For this purpose, we propose the model Lagrangian, which is in the first-order and singular and contains two-kind of Chern-Simons like terms.  Starting with this Langrangian, we shall construct the noncommutative quantum Hamiltonian system. 

\subsection{Noncommutativity Matrix $\Theta$, $\Xi$}

\ts{12pt}Let $\Theta$ and $\Xi$ be the totally antisymmetric matrices defined as follows:  
$$
\Theta=\theta\varepsilon,\hs{48pt}\Xi=\eta\varepsilon,
\eqno{(3.1)}
$$
where $\theta$ is the constant parameter describing the noncommutativity of coordinates and $\eta$, that of momenta, and
$\varepsilon$ is the completely antisymmetric tensor defined as
$$
\varepsilon^{ij}=1\hs{12pt}(i>j), \hs{24pt}\varepsilon^{ji}=-\varepsilon^{ij}\hs{36pt}(i,j=1,\cdots,N).
\eqno{(3.2)}
$$
These matrices satisfy 
$$
\Theta\Xi=\Xi\Theta,\hs{48pt}(\Theta\Xi)^t=\Theta\Xi.
\eqno{(3.3)}
$$
In terms of $\Theta$ and $\Xi$, then, the following matrices are defined:
$$
\begin{array}{cll}
\ts{114pt}&G=\Theta\Xi=\Xi\Theta,&\hs{186pt}(3.4a)\vs{6pt}\\
&M=I+\dps{\f14}G,&\hs{186pt}(3.4b)\vs{6pt}\\
&\bM=I-\dps{\f14}G,&\hs{186pt}(3.4c)
\end{array}
$$
which are symmetric and commutable with $\Theta$, $\Xi$, and therefore become commutable with each other:
$$
\begin{array}{cll}
\ts{54pt}&G^t=G,\hs{12pt}M^t=M,\hs{12pt}\bM^t=\bM,&\hs{57pt}(3.5a)\vs{12pt}\\
&G\Theta=\Theta G,\hs{12pt}G\Xi=\Xi G,&\hs{57pt}(3.5b)\vs{12pt}\\
&M\Theta=\Theta M,\hs{6pt}M\Xi=\Xi M,\hs{6pt}\bM\Theta=\Theta\bM,\hs{6pt}\bM\Xi=\Xi\bM.&\hs{57pt}(3.5c)
\end{array}
$$
\\
\ts{12pt}Due to Eqs.(3.4), there exist the inverses $M^{-1}$ and $\bM^{-1}$, which also satisfy the same properties as $M$, $\bM$.

\subsection{Noncommutative Quantum System}

\subsubsection{Primary Hamiltonian System}

\ts{12pt}Consider the dynamical system described by the first-order singular Lagrangian $L$ 
$$
\begin{array}{rcl}
L&=&L(x,\dot{x},v,\dot{v},u,\dot{u},\lambda,\dot{\lambda})\vs{6pt}\\
&=&\dps{\dot{x}^i\bM_{ij}v_j-\lambda_i(u^i-x^i)-\f12\dot{v}_i\Theta^{ij}v_j-\f12\dot{u}^i\Xi_{ij}u^j-h_0(x,v,u)},
\end{array}
\eqno{(3.6)}
$$
where $h_0(x,v,u)$ corresponds to the Hamiltonian in the final constraint quantum system $\mcS^*$.\\
\ts{12pt}Following the canonical quantization formulation for constraint systems\cite{MN1,Dirac}, then, the initial unconstraint quantum system $\mcS=(\mcC,\mcA(\mcC),H(\mcC),\mcK)$ is obtained as follows: 
$$
\ts{-72pt}\mcC=\{(x^i,p^x_i),(v_i,\pi_v^i),(u^i,\pi^u_i),(\lambda_i,\pi^i_{\lambda})|i=1,\cdots,N\},
\eqno{(3.7a)}
$$

$$
\ts{-60pt}\begin{array}{lcl}
\mcA(\mcC)&:&\commut{x^i}{p^x_j}=i\hbar\delta^i_j,\commut{v_i}{\pi_v^j}=i\hbar\delta_i^j,\commut{u^i}{p^u_j}=i\hbar\delta^i_j,
\vs{6pt}\\
& &\commut{\lambda_i}{\pi_{\lambda}^j}=i\hbar\delta_i^j,\q\mbox{(the others)}=0,
\end{array}
\eqno{(3.7b)}
$$

$$
\ts{-126pt}H=\sum^4_{n=1}\{\mu^i_{(n)},\phi^{(n)}_i\}+\{\lambda_i,\psi^{\mbtn{(1)}}_i\}+h_{0}(x,v,u),
\eqno{(3.7c)}
$$

$$
\begin{array}{lcl}
\mcK&=&\{\phi^{\mbtn{(1)}}_i,\phi^{\mbtn{(2)}}_i,\phi^{\mbtn{(3)}}_i,\phi^{\mbtn{(4)}}_i,\psi^{\mbtn{(1)}}_i,\psi^{\mbtn{(2)}}_i|i=1,\cdots,N\}\vs{6pt}\\
& &\mbox{with}\\
& &\begin{array}{ll}
\phi^{\mbtn{(1)}}_i=\bM_{ij}v_j-p^x_i,&\phi^{\mbtn{(2)}}_i=\pi^i_v+\dps{\f12}\Theta^{ij}v_j,\\
\phi^{\mbtn{(3)}}_i=p^u_i+\dps{\f12}\Xi_{ij}u^j, &\phi^{\mbtn{(4)}}_i=\pi^i_{\lambda},\\
\psi^{\mbtn{(1)}}_i=u^i-x^i, &\psi^{\mbtn{(2)}}_i=\lambda_i-(W^{-1})_{ij}\mathcal{H}^{(0)}_j(x,v,u).
\end{array}
\end{array}
\eqno{(3.7d)}
$$
Here, $\phi^{\mbtn{(n)}}(n=1,\cdots,4)$ are the constraint-operators corresponding to the primary constraints $\phi^{\mbtn{(n)}}\approx 0$ due to 
the singularity of the Lagrangian (3.6), $\psi^{\mbtn{(n)}}(n=1,2)$, those corresponding to the secondary ones $\psi^{\mbtn{(n)}}\approx 0$, and
$$
W=I+\bM^{-1}G\bM^{-1},
\eqno{(3.8)}
$$ 
$$
\mathcal{H}^{(0)}_i(x,v,u)=((\bM^{-1}G\bM^{-1})_{ij}\partial^x_j+(\bM^{-1}\Xi)_{ij}\partial^j_v-\partial^u_i)h_0(x,v,u).
\eqno{(3.9)}
$$
The commutator algebra $\mcA(\mcK)$ is presented in Appendix A. 
The Lagrange multiplier operators $\mu^i_{\mbtn{(n)}}(n=1,\cdots,4)$ are determined together with the secondary constraints through the consistency conditions for the time evolusion of the constraint-operators (see Appendix B).

\subsubsection{Successive Projections of $\mcS$}

\ts{12pt}According to the structure of the commutator algebra (A.1), it is convenient to classify $\mcK$ into the following three subsets  :
$$
\mcK=\mcK^{(\mbtn{A})}\oplus\mcK^{\mbtn{(B)}}\oplus\mcK^{\mbtn{(C)}}
\eqno{(3.10a)}
$$
with
$$
\mcK^{(\mbtn{A})}=\{\phi^{\mbtn{(1)}},\phi^{\mbtn{(2)}}\},\hs{12pt}\mcK^{\mbtn{(B)}}=\{\phi^{\mbtn{(3)}},\psi^{\mbtn{(1)}}\},\hs{12pt}\mcK^{\mbtn{(C)}}=\{\phi^{\mbtn{(4)}},\psi^{\mbtn{(2)}}\}.
\eqno{(3.10b)}
$$
As well as the Dirac bracket formalism, the POM satisfies the {\it iterative} property\cite{HRT,MN3}.\\ 
\ts{12pt}Starting with the initial system (3.7), we shall construct the constraint quantum system $\mcS^*$ through the successive operations of projection operators\cite{MN4}. For this purpose, we first rearrange the subsets (3.10b) to $\mcK^{(n)}\e (n=1,2,3)$, and let $\Phat^{(n)}$ be the projection operator aasociated to the subset $\mcK^{(n)}$, that is, $\Phat^{(n)}\mcK^{(n)}=0$. Then, the successive projections of the operators of the system by $\Phat^{(n)}$ $(n=1,2,3)$ can be carried out through the program designated by the following diagram :
$$
\mcC\stackrel{\Phat^{(1)}}{\longrightarrow}\mcC^{(1)}\stackrel{\Phat^{(2)}}{\longrightarrow}\mcC^{(2)}\stackrel{\Phat^{(3)}}\longrightarrow\mcC^{(3)},
\eqno{(3.11)}
$$
where 
$$
\mcC^{(n)}=\Phat^{(n)}\mcC^{(n-1)}\hs{60pt}(n=1,2,3)
\eqno{(3.12a)}
$$
with $\mcC^{(0)}=\mcC$, which satisfy 
$$
\mcK^{(n)}(\mcC^{(n)})=0.
\eqno{(3.12b)}
$$
Then, $Z^{(n)}$ for the subsets $\mcK^{(n)}$ $(n=1,2,3)$ consist of the operators in $\mcC^{(n-1)}$,
$$
Z^{(n)}=Z^{(n)}(\mcC^{(n-1)}).
\eqno{(3.13)}
$$
From (2.7), therefore, the projection operators $\Phat^{(n)}$ are also represented as 
$$
\Phat^{(n)}=\Phat^{(n)}(\mcC^{
(n-1)})\hs{60pt}(n=1,2,3).
\eqno{(3.14)}
$$

\subsubsection{Successive projection I}

\ts{12pt}
Let $\mcK^{(n)}$ $(n=1,2,3)$ be $\mcK^{(1)}=\mcK^{\mbtn{(C)}}$, $\mcK^{(2)}=\mcK^{\mbtn{(B)}}$ and $\mcK^{(3)}=\mcK^{\mbtn{(A)}}$. Then, we shall accomplish the successive projection of operators through the following diagram:
$$ 
\mbox{I}\ : \ \Phat^{(1)}\mcK^{\mbtn{(C)}}=0\longrightarrow\Phat^{(2)}\mcK^{\mbtn{(B)}}=0\longrightarrow\Phat^{(3)}\mcK^{\mbtn{(A)}}=0.
\eqno{(3.15)}
$$
The ACCS $Z^{\mbtn{(n)}}_{\alpha}$ of the projection operators $\Phat^{(n)}$ ($n=1,2,3$) are given as follows, respectively: 
$$
\begin{array}{clcl}

(1)&Z^{\mbtn{(1)}}_{\alpha}=Z^{\mbtn{(1)}}_{\alpha}(\mcC)&=&\left\{\begin{array}{rcll}\xi^{\mbtn{(1)}}_i&=&\psi^{\mbtn{(2)}}_i&\hs{114pt}(\alpha=i),\vs{6pt}\\
\pi^{\mbtn{(1)}}_i&=&\phi^{\mbtn{(4)}}_i&\hs{114pt}(\alpha=i+N),
\end{array}\right.
\vs{12pt}\\

(2)&Z^{\mbtn{(2)}}_{\alpha}=Z^{\mbtn{(2)}}_{\alpha}(\mcC^{(1)})&=&\left\{\begin{array}{rcll}\xi_i^{\mbtn{(2)}}&=&\psi^{\mbtn{(1)}}_i&\hs{60pt}(\alpha=i),\vs{6pt}\\
\pi_i^{\mbtn{(2)}}&=&\phi^{\mbtn{(3)}}_i-\dps{\f12}\Xi_{ij}\psi^{\mbtn{(1)}}_j&\hs{60pt}(\alpha=i+N),
\end{array}\right.
\vs{12pt}\\

(3)&Z^{\mbtn{(3)}}_{\alpha}=Z^{\mbtn{(3)}}_{\alpha}(\mcC^{(2)})&=&\left\{\begin{array}{rcll}\xi_i^{\mbtn{(3)}}&=&(M^{-1})_{ij}(\phi^{\mbtn{(1)}}_j+\dps{\f12}\Xi_{jk}\phi^{\mbtn{(2)}}_k)&\hs{9pt}(\alpha=i),\vs{12pt}\\
\pi_i^{\mbtn{(3)}}&=&(M^{-1})_{ij}(\phi^{\mbtn{(2)}}_j-\dps{\f12}\Theta^{jk}\phi^{\mbtn{(1)}}_k)&\hs{9pt}(\alpha=i+N),
\end{array}\right.
\end{array}
\eqno{(3.16)}
$$
$$
\hs{60pt}(\alpha=1,\cdots,2N\q;\q i=1,\cdots,N).
$$
\ts{12pt}Let $\Phat$ be $\Phat=\Phat^{\mbtn{(3)}}\Phat^{\mbtn{(2)}}\Phat^{\mbtn{(1)}}$, then, $\mcC^{(3)}$ is obtained as follows:
$$
\begin{array}{rcl}
\mcC^{\mbtn{(3)}}&=&\Phat\mcC=\mcC\{(\Phat x,\Phat p^x),(\Phat v,\Phat \pi_v),(\Phat  u,\Phat p^u),(\Phat\lambda,\Phat\pi_{\lambda})\}\vs{6pt}\\
&=&\mcC^{\mbtn{(3)}}\{(x,p^x),(v,\pi_v),(u,p^u)\}
\end{array}
\eqno{(3.17a)}
$$
with
$$
\begin{array}{rcl}
\lambda_i&=&(W^{-1})_{ij}\Phat^{\mbtn{(3)}}\mcH^{(0)}_j(x,v,u),\vs{6pt}\\
\pi^i_{\lambda}&=&0.
\end{array}
\eqno{(3.17b)}
$$
\ts{12pt}Under the operation of $\Phat^{(3)}$ in the process I, now, the operators $x$,$v$ and $u$  become noncommutable with each other. For any operator $O(x,v,u)$, therefore, the projection of $O(x,v,u)$ by $\Phat^{(3)}$ would not always be equivalent to the operator $O$ consisiting of the projections of $x,v,u$, that is,
$$
\Phat^{(3)}O(x,v,u)\neq O(\Phat^{(3)}x,\Phat^{(3)}v,\Phat^{(3)}u).
\eqno{(3.18)}
$$
The projection of $\mcH^{(0)}_j(x,v,u)$ in Eq.(3.17b) is thus denoted as the form of $\Phat^{(3)}\mcH^{(0)}_j(x,v,u)$ .\\
\ts{12pt}From Eqs.(2.11a), (3.16), the {\it hyper}-operators $\Qhat^{(n)}_{\eta\zeta}$ for $\Phat^{(n)}$ $(n=1,2,3)$ are described by 
$$
\Qhat^{(n)}_{\eta\zeta}=\hat{\xi}^{(n)(-)}_i(\eta)\hat{\pi}^{(n)(-)}_i(\zeta)-\hat{\pi}^{(n)(-)}_i(\eta)\hat{\xi}^{(n)(-)}_i(\zeta)\hs{36pt}(n=1,2,3),
\eqno{(3.19)}
$$
the explisit forms of which are presented in Apppndix C.\\
\ts{12pt}Using the commutator formulas and the symmetrized ones (2.12) and (2.13), now, one obtains the commutator algebra $\mcA(\mcC^{(3)})$, which is presented in Appendix D.  
From the structure of $\mcA(\mcC^{(3)})$, thus, we shall adopt $\{(x,v)\}$ as the {\it projected} CCS $\mcC^*$ in $\mcS^*$:
$$
\mcC^*=\{(x^i,v_i)|i=1,\cdots,N\}=\{(x,v)\}
\eqno{(3.20)}
$$
with
$$
\begin{array}{rclcrcl}
u^i&=&x^i,& &\lambda_i&=&(W^{-1})_{ij}\Phat^{\mbtn{(III)}}\mcH^{(0)}_j(x,v,u(x)),\vs{12pt}\\
p^x_i&=&\bM_{ij}v_j,& &\pi^v_i&=&-\dps{\f12}\Theta_{ij}v_j,\vs{12pt}\\
p^u_i&=&-\dps{\f12}\Xi_{ij}x^j,& &\pi^i_{\lambda}&=&0.
\end{array}
\eqno{(3.21)}
$$
Then, the canonical structure $\mcA^*(\mcC^*)$ is represented as
$$
\begin{array}{lcrcl}
\mcA^*(\mcC^*)&:&\commut{x^i}{x^j}&=&i\hbar(M^{-1}\Theta M^{-1})_{ij},\vs{12pt}\\
& &\commut{x^i}{v_j}&=&i\hbar(M^{-1}\bar{M}M^{-1})_{ij},\vs{12pt}\\
& &\commut{v_i}{v_j}&=&i\hbar(M^{-1}\Xi M^{-1})_{ij},
\end{array}
\eqno{(3.22)}
$$
and the projected Hamiltonian $H^*$ becomes
$$
H^*=\Phat^{(3)}h_0(x,v,u(x)),
\eqno{(3.23)}
$$
where $h_0(x,v,u(x))\in \mcC^{\mbtn{(2)}}$. 
\vs{6pt}\\
\ts{12pt}Thus, we have obtained the constraint quantum system in the process I,
$$
\mcS^*=(\mcC^*,\mcA^*(\mcC^*),H^*(\mcC^*)).
\eqno{(3.24)}
$$

\subsubsection{Successive projection II}

\ts{12pt}The constraint operators $\phi^{\mbtn{(4)}}_i=\pi_{\lambda}^i$ in $\mcK^{\mbtn{(C)}}$ are commutable with the canonical pairs $(x,p^x),(v,\pi_v)$ and $(u,p^u)$. Therefore, it turns out that the result of the successive projection of the system does not depend on the order of the operation of $\Phat^{\mbtn{(C)}}$. \\
\ts{12pt}Consider, then, the projection process where the subsets $\mcK^{\mbtn{(A)}}$ and $\mcK^{\mbtn{(B)}}$ are interchanged in the process (3.15),
$$
\mbox{II}\ :\ \Phat^{(1)}\mcK^{\mbtn{(C)}}=0\longrightarrow\Phat^{(2)}\mcK^{\mbtn{(A)}}=0\longrightarrow\Phat^{(3)}\mcK^{\mbtn{(B)}}.
\eqno{(3.25)}
$$
\ts{12pt}The ACCS's in the process (3.25) are given as follows:

$$
\begin{array}{clcl}

(1)&Z^{\mbtn{(1)}}_{\alpha}=Z^{\mbtn{(1)}}_{\alpha}(\mcC)&=&\left\{\begin{array}{rcll}\xi^{\mbtn{(1)}}_i&=&\psi^{\mbtn{(2)}}_i&\hs{136pt}(\alpha=i),\vs{6pt}\\
\pi^{\mbtn{(1)}}_i&=&\phi^{\mbtn{(4)}}_i&\hs{138pt}(\alpha=i+N),
\end{array}\right.
\vs{12pt}\\

(2)&Z^{\mbtn{(2)}}_{\alpha}=Z^{\mbtn{(2)}}_{\alpha}(\mcC^{(1)})&=&\left\{\begin{array}{rcll}\xi_i^{\mbtn{(2)}}&=&(\bM^{-1})_{ij}\phi^{\mbtn{(1)}}_j&\hs{48pt}(\alpha=i),\vs{6pt}\\
\pi_i^{\mbtn{(2)}}&=&\phi^{\mbtn{(2)}}_i-\dps{\f12}(\bM^{-1}\Theta)_{ij}\phi^{\mbtn{(1)}}_j&\hs{48pt}(\alpha=i+N),
\end{array}\right.
\vs{12pt}\\

(3)&Z^{\mbtn{(3)}}_{\alpha}=Z^{\mbtn{(3)}}_{\alpha}(\mcC^{(2)})&=&\left\{\begin{array}{rcll}\xi_i^{\mbtn{(3)}}&=&(M^{-1})_{ij}(\bM_{jk}\psi^{\mbtn{(1)}}_k+\dps{\f12}\Xi_{jk}\phi^{\mbtn{(3)}}_k)&\hs{9pt}(\alpha=i),\vs{12pt}\\
\pi_i^{\mbtn{(3)}}&=&(M^{-1})_{ij}(\phi^{\mbtn{(3)}}_j-\dps{\f12}(\Xi\bM)_{jk}\psi^{\mbtn{(1)}}_k)&\hs{9pt}(\alpha=i+N),
\end{array}\right.
\end{array}
\eqno{(3.26)}
$$
and the {\it hyper}-operators $\Qhat^{(n)}_{\eta\zeta}$   $(n=1,2,3)$ for the process II is also presented in Appendix C.\\
\ts{12pt}As well as in the case of the successive projection I, then, we obtain the constraint quantum system $\mcS^*$ in II, which is identical with $\mcS^*$ in I, except that the projection of an operator $O(x,v,u)$ is represented as $\Phat^{(3)}\Phat^{(2)}O(x,v,u)$ in II. \\
\ts{12pt}Consequently, one can express the projected Hamiltonian $H^*(\mcC^*)$ with the unified form as follows:
$$
H^*(\mcC^*)=H^*(x,v)=\left\{\begin{array}{ll}\Phat^{(3)}h_0(x,v,u(x))&\hs{48pt}\mbox{I}\vs{6pt}\\
\Phat^{(3)}\Phat^{(2)}h_0(x,v,u)&\hs{48pt}\mbox{II}.
\end{array}
\right.
\eqno{(3.27)}
$$ 

\section{{\it Exact} Canonically Conjugate Set}

\ts{12pt}The commutator algebra in the constraint quantum system $\mcS^*$ has been given by Eq.(3.22). In order to construct the {\it exact} CCS in $\mcS^*$, we first introduce 
$$
q^i=M_{ij}x^j,\hs{48pt}p_i=M_{ij}v_j,
\eqno{(4.1)}
$$ 
which obey the commutator algebra $\mcA(q,p)$,
$$
\begin{array}{rcr}
\mcA(q,p)\ :\ \commut{q^i}{q^j}&=&i\hbar\Theta^{ij},\vs{6pt}\\
\commut{q^i}{p_j}&=&i\hbar\bM_{ij},\vs{6pt}\\
\commut{p_i}{p_j}&=&i\hbar\Xi_{ij}.
\end{array}
\eqno{(4.2)}
$$
\ts{12pt}Let $\mcC^*(Q,P)$ be the {\it exact} CCS with the commutator algebra 
$$
\commut{Q^i}{Q^j}=0,\hs{36pt}\commut{Q^i}{P_j}=i\hbar\delta^i_j,\hs{36pt}\commut{P_i}{P_j}=0,
\eqno{(4.3)}
$$ 
we next introduce the projection operators defined by
$$
\begin{array}{l}
\dps{\Phat_{\mbtn{Q}}=\sum^{\infty}_{n=0}\f1{n!}\hat{Q}^{(+)n}\hat{P}^{(-)n}},\vs{12pt}\\
\dps{\Phat_{\mbtn{P}}=\sum^{\infty}_{n=0}\f{(-1)^n}{n!}\hat{P}^{(+)n}\hat{Q}^{(-)n}},
\end{array}
\eqno{(4.4)}
$$
which satisfy
$$
\Phat_{\mbtn{Q}}Q^i=\Phat_{\mbtn{P}}P_i=0\hs{48pt}(i=1,\cdots,N).
\eqno{(4.5)}
$$  
\ts{12pt}Following the POM\cite{MN1,MN5}, then, the {\it exact} CCS is obtained in the following way: 
$$
Q^i=\Phat_{\mbtn{P}}q^i=\sum^{\infty}_{n=0}\f{(-1)^n}{n!}\hat{P}^{(+)n}\hat{Q}^{(-)n}q^i=\sum^{\infty}_{n=0}\f{(-1)^n}{(n+1)!}\hat{P}^{(+)n}\hat{q}^{(-)n}q^i,
\eqno{(4.6a)}
$$
$$
P_i=\Phat_{\mbtn{Q}}p_i=\sum^{\infty}_{n=0}\f1{n!}\hat{Q}^{(+)n}\hat{P}^{(-)n}p_i=\sum^{\infty}_{n=0}\f1{(n+1)!}\hat{Q}^{(+)n}\hat{p}^{(-)n}p_i.
\eqno{(4.6b)}
$$ 
From the commutator algebra (4.2), Eqs.(4.6) become
$$
Q^i=q^i+\f12\Theta^{ij}P_j,
\eqno{(4.7a)}
$$
$$
P_i=p_i-\f12\Xi_{ij}Q^j.
\eqno{(4.7b)}
$$
Then, one immediately obtains 
$$
Q^i=M^{-1}_{ij}(q^j+\f12\Theta^{jk}p_k)=x^i+\f12\Theta^{ij}v_j,
\eqno{(4.8a)}
$$ 
$$
P_i=M^{-1}_{ij}(p_j-\f12\Xi_{jk}q^k)=v_i-\f12\Xi_{ij}x^j,
\eqno{(4.8b)}
$$
which are equivalent to the formulas in the so-called Bopp shift\cite{Bopp1}. \\
\ts{12pt}According to Eqs.(4.8), $x^i$ and $v_i$ become
$$
x^i=(M^{-1})_{ij}(Q^j-\f12\Theta^{jk}P_k),\hs{36pt}v_i=(M^{-1})_{ij}(P_j+\f12\Xi_{jk}Q^k).
\eqno{(4.9)}
$$
The projected Hamiltonian $H^*$ is thus represented as follows:
$$
H^*=
H^*(x,v)=H^*(M^{-1}(Q-\f12\Theta P),M^{-1}(P+\f12\Xi Q))=\tilde{H}^*(QP).
\eqno{(4.10)}
$$ 
which will be shown to contain the quantum corrections due to the noncommutativity among the {\it projected} CCS and the ACCS\cite{MN6}.\\
\ts{12pt}The projected Hamiltonian $H^*$ is represented also in terms of $(q,\ p)$ defined by Eq.(4.1) as follows:
$$
H^*=H^*(M^{-1}q,M^{-1}p)=\mathcal{H}^*(q,p).
\eqno{(4.11)}
$$
Then, $\mcS^*$ becomes 
$$
\mcS^*=(\mcC^*(q,p),\mcA^*(\mcC^*),\mathcal{H}^*(q,p)),
\eqno{(4.12)}
$$ 
where $\mcA^*(\mcC^*)$ is given by Eq.(4.2). 

\section{Discussion and Conclusions}

\ts{12pt}Starting with the first-order singular Lagrangian , we have shown that the noncommutative quantum system $\mcS^*$ is exactly constructed through the POM with the star-product quantization. The canonical structure in $\mcS^*$ has been defined by the commutator-algebra $\mcA^*(x,v)$, which contains all-order of the noncommutativity parameters through $M^{-1}$. \\
\ts{12pt}We have shown that the canonical structure $\mcA^*$ is represented also in terms of the transformed CCS, $\mcC^*(q,p)$, which obeys the ordinaly type (4.1) of commutator algebra. Using $\mcC^*(q,p)$, further, we have constructed the {\it exact} CCS $\mcC^*(Q,P)$.\\
\ts{12pt}We finally discuss the alternative model Lagrangian to realize both of space-space and momentum-momentum noncommtativities, which is proposed by
$$
\begin{array}{rcl}
L'&=&L'(x,\dot{x},v,\dot{v},u,\dot{u},\lambda,\dot{\lambda})\vs{6pt}\\
&=&\dps{\dot{x}^iv_i-\lambda_i(\bar{M}_{ij}u^j-x^i)-\f12\dot{v}_i\Theta^{ij}v_j-\f12\dot{u}^i\Xi_{ij}u^j-h_0(x,v,u)}.
\end{array}
\eqno{(5.1)}
$$
Let $p^i_x=\partial L'/\partial \dot{x}^i$, then, we will obtain the noncommutative quantum system $\mcS^*=(\mcC^*(u,p_x),H^*(u,p_x))$, of which canonical structure $\mcA^*(u,p_x)$ is equivalent to $\mcA^*(x,v)$ defined by (3.22). Furthermore, the noncommutative quantum system constructed in Appendix E also holds the canonical structure equivalent to (3.22). \\
\ts{12pt} The projected Hamiltonians contain the quantum corrections due to the noncommutativity among the {\it projected} CCS and the ACCS. This problem will be investigated in near future.

\newpage

\appendix

\ts{-18pt}{\bf \Large Appendix}

\section{Commutator Algebra of Constraint operators}
\ts{12pt}Under the commutator algebra $\mcA(\mcC)$, the commutation relations among the initial constraint-operators are given by 
$$
\begin{array}{lcl}
\mcA(\mcK):& &\vs{6pt}\\
\commut{\phi^{\mbtn{(1)}}_i}{\phi^{\mbtn{(2)}}_j}=i\hbar\bM_{ij},& &
\commut{\phi^{\mbtn{(1)}}_i}{\psi^{\mbtn{(1)}}_j}=-i\hbar\delta_{ij},\vs{12pt}\\
\commut{\phi^{\mbtn{(1)}}_i}{\psi^{\mbtn{(2)}}_j}=-i\hbar(W^{-1})_{jk}\partial^x_i\mcH^{(0)}_k(x,v,u),& &
\commut{\phi^{\mbtn{(2)}}_i}{\phi^{\mbtn{(2)}}_i}=i\hbar\Theta^{ij},\vs{12pt}\\
\commut{\phi^{\mbtn{(2)}}_i}{\psi^{\mbtn{(2)}}_j}=i\hbar(W^{-1})_{jk}\partial^i_v\mcH^{(0)}_k(x,v,u),& &
\commut{\phi^{\mbtn{(3)}}_i}{\phi^{\mbtn{(3)}}_j}=i\hbar\Xi_{ij},\vs{12pt}\\
\commut{\phi^{\mbtn{(3)}}_i}{\psi^{\mbtn{(1)}}_j}=-i\hbar\delta_{ij},& &
\commut{\phi^{\mbtn{(3)}}_i}{\psi^{\mbtn{(2)}}_j}=i\hbar(W^{-1})_{jl}\partial^u_i\mathcal{H}^{(0)}_l,\vs{12pt}\\
\commut{\phi^{\mbtn{(4)}}_i}{\psi^{\mbtn{(2)}}_j}=-i\hbar\delta_{ij},
& &
\mbox{(the others)}=0.
\end{array}
\eqno{(\mbox{A}1)}
$$

\section{Lagrange multiplier operators}

\ts{12pt}The Lagrange muliplier operators in the Hamiltonian (3.7c) are given as follow:
$$
\begin{array}{rclcl}
\mu^i_{\mbtn{(1)}}&=&\mu^i_{\mbtn{(1)}}(x,v,u,\lambda)&=&-(G^{-1}\Theta)_{ij}(\lambda_j+\partial^u_jh_0(x,v,u)),\vs{12pt}\\
\mu^i_{\mbtn{(2)}}&=&\mu^i_{\mbtn{(2)}}(x,v,u,\lambda)&=&(\bM^{-1})_{ij}(\lambda_j-\partial^x_jh_0(x,v,u)),\vs{12pt}\\
\mu^i_{\mbtn{(3)}}&=&\mu^i_{\mbtn{(3)}}(x,v,u,\lambda)&=&(G^{-1}\Theta)_{ij}(\lambda_j+\partial^u_jh_0(x,v,u)),\vs{12pt}\\
\mu^i_{\mbtn{(4)}}&=&\mu^i_{\mbtn{(4)}}(x,v,u,\lambda)&=&(W^{-1})_{ij}(-\mu^k_{\mbtn{(1)}}\partial^x_k+\mu^k_{\mbtn{(2)}}\partial^k_v+\mu^k_{\mbtn{(3)}}\partial^u_k)\mathcal{H}^0_j(x,v,u).
\end{array}
\eqno{(\mbox{B}1)}
$$

\newpage

\section{The representation of $\Qhat_{\eta\zeta}$}

The explisit forms of {\it hyper}-operators $\Qhat^{(n)}_{\eta\zeta}$ $(n=1,2,3)$ in the projection processes I and II.
\subsection{Projection process I}
$$
\begin{array}{rcl}
\Qhat^{(1)}_{\eta\zeta}&=&\hat{\psi}^{(2)(-)}_k(\eta)\hat{\pi}^{k(-)}_{\lambda}(\zeta)-\hat{\pi}^{k(-)}_{\lambda}(\eta)\hat{\psi}^{(2)(-)}_k(\zeta)\vs{12pt}\\
& &\hs{48pt}\mbox{with}\hs{24pt}\psi^{(2)}_k=\lambda_k-(W^{-1})_{kl}\mcH^{(0)}_l(x,v,u),
\vs{12pt}\\
\Qhat^{(2)}_{\eta\zeta}&=&-\Xi_{kl}\hat{x}^{k(-)}(\eta)\hat{x}^{l(-)}(\zeta)\vs{12pt}\\
& &+(\hat{u}^{k(-)}(\eta)\hat{p}^{u(-)}_k(\zeta)-\hat{p}^{u(-)}_k(\eta)\hat{u}^{k(-)}(\zeta))\vs{12pt}\\
& &+(\hat{p}^{u(-)}_k(\eta)\hat{x}^{k(-)}(\zeta)-\hat{x}^{k(-)}(\eta)\hat{p}^{u(-)}_k(\zeta))\vs{12pt}\\
& &+\dps{\f12}\Xi_{kl}(\hat{x}^{k(-)}(\eta)\hat{u}^{l(-)}(\zeta)+\hat{u}^{k(-)}(\eta)\hat{x}^{l(-)}(\zeta)),\vs{12pt}\\
\Qhat^{(3)}_{\eta\zeta}&=&-(M^{-1}\Theta M^{-1})_{kl}\hat{p}^{x(-)}_k(\eta)\hat{p}^{x(-)}_l(\zeta)\vs{12pt}\\
& &-(M^{-1}\Xi M^{-1})_{kl}\hat{\pi}^{k(-)}_v(\eta)\hat{\pi}^{l(-)}_v(\zeta)\vs{12pt}\\
& &+\dps{\f14}(M^{-1}G\Theta M^{-1})_{kl}\hat{v}^{(-)}_k(\eta)\hat{v}^{(-)}_l(\zeta)\vs{12pt}\\
& &+(M^{-1}\bM M^{-1})_{kl}(\hat{\pi}^{k(-)}_v(\eta)\hat{p}^{x(-)}_l(\zeta)-\hat{p}^{x(-)}_k(\eta)\hat{\pi}^{l(-)}_v(\zeta))\vs{12pt}\\
& &+\dps{\f12}(M^{-1}\Theta\bM M^{-1})_{kl}(\hat{p}^{x(-)}_k(\eta)\hat{v}^{(-)}_l(\zeta)+\hat{v}^{(-)}_k(\eta)\hat{p}^{x(-)}_l(\zeta))\vs{12pt}\\
& &+(M^{-1}(I+\dps{\f1{16}}G^2)M^{-1})_{kl}(\hat{v}^{(-)}_k(\eta)\hat{\pi}^{l(-)}_v(\zeta)-\hat{\pi}^{k(-)}_v(\eta)\hat{v}^{(-)}_l(\zeta)).
\end{array}
\eqno{(\mbox{C}1)}
$$
\subsection{ Projecion process II}
$$
\begin{array}{rcl}
\Qhat^{(1)}_{\eta\zeta}&=&\hat{\psi}^{(2)(-)}_k(\eta)\hat{\pi}^{k(-)}_{\lambda}(\zeta)-\hat{\pi}^{k(-)}_{\lambda}(\eta)\hat{\psi}^{(2)(-)}_k(\zeta)\vs{18pt}\\
& &\hs{48pt}\mbox{with}\hs{24pt}\psi^{(2)}_k=\lambda_k-(W^{-1})_{kl}\mcH^{(0)}_l(x,v,u),
\vs{12pt}\\

\Qhat^{(2)}_{\eta\zeta}&=&
-(\bM^{-1}\Theta \bM^{-1})_{kl}\hat{p}^{x(-)}_k(\eta)\hat{p}^{x(-)}_l(\zeta)\vs{12pt}\\& &-(\bM^{-1})_{kl}(\hat{p}^{x(-)}_k(\eta)\hat{\pi}^{l(-)}_v(\zeta)-\hat{\pi}^{k(-)}_v(\eta)\hat{p}^{x(-)}_l(\zeta))\vs{12pt}\\
& &+(\hat{v}^{(-)}_k(\eta)\hat{\pi}^{(-)k}_v(\zeta)-\hat{\pi}^{(-)k}_v(\eta)\hat{v}^{(-)}_k(\zeta)\vs{12pt}\\
& &+\dps{\f12}(\bM^{-1}\Theta)_{kl}(\hat{p}^{x(-)}_k(\eta)\hat{v}^{(-)}_l(\zeta)+\hat{v}^{(-)}_k(\eta)\hat{p}^{x(-)}_l(\zeta))\vs{12pt}\\

\Qhat^{(3)}_{\eta\zeta}&=&-(M^{-1}\bM\Xi \bM M^{-1})_{kl}\hat{x}^{k(-)}_v(\eta)\hat{x}^{l(-)}_v(\zeta)\vs{12pt}\\
& &-(M^{-1}\Theta M^{-1})_{kl}\hat{p}^{u(-)}_k(\eta)\hat{p}^{u(-)}_l(\zeta)\vs{12pt}\\
& &+\dps{\f14}(M^{-1}G\Xi M^{-1})_{kl}\hat{u}^{k(-)}(\eta)\hat{u}^{l(-)}(\zeta)\vs{12pt}\\
& &+(M^{-1}\bM^2 M^{-1})_{kl}(\hat{p}^{u(-)}_k(\eta)\hat{x}^{l(-)}(\zeta)-\hat{x}^{k(-)}(\eta)\hat{p}^{u(-)}_l(\zeta))\vs{12pt}\\
& &+\dps{\f12}(M^{-1}\bM\Xi\bM M^{-1})_{kl}(\hat{x}^{k(-)}(\eta)\hat{u}^{l(-)}(\zeta)+\hat{u}^{k(-)}(\eta)\hat{x}^{l(-)}(\zeta))\vs{12pt}\\
& &+(M^{-1}(I+\dps{\f1{16}}G^2)M^{-1})_{kl}(\hat{u}^{k(-)}(\eta)\hat{p}^{u(-)}_l(\zeta)-\hat{p}^{u(-)}_k(\eta)\hat{u}^{l(-)}(\zeta)).
\end{array}
\eqno{(\mbox{C}2)}
$$

\newpage

\section{Commutator Algebra $\mcA(\mcC^{(3)})$}

$$
\begin{array}{rclcrcl}

\commut{x^i}{x^j}&=&i\hbar(M^{-1}\Theta  M^{-1})_{ij},&\hs{18pt}&\commut{x^i}{u^j}&=&i\hbar(M^{-1}\Theta M^{-1})_{ij},
\vs{12pt}\\

\commut{x^i}{p^x_j}&=&i\hbar(M^{-1}\bar{M}\bar{M}M^{-1})_{ij},&\hs{18pt}&\commut{x^i}{p^u_j}&=&i\hbar\dps{\f12}(M^{-1}GM^{-1})_{ij},
\vs{12pt}\\

\commut{p^x_i}{p^x_j}&=&i\hbar(M^{-1}\bM\Xi \bM M^{-1})_{ij},&\hs{18pt}&\commut{v^i}{p^x_j}&=&i\hbar(M^{-1}\Xi\bM M^{-1})_{ij},
\vs{12pt}\\

\commut{v_i}{v_j}&=&i\hbar(M^{-1}\Xi M^{-1})_{ij},&\hs{18pt}&\commut{v_i}{u^j}&=&-i\hbar(M^{-1}\bM M^{-1})_{ij},
\vs{12pt}\\

\commut{v_i}{\pi_j^v}&=&i\hbar\dps{\f12}(M^{-1}GM^{-1})_{ij},&\hs{18pt}&\commut{v_i}{p^u_j}&=&-i\hbar\dps{\f12}(M^{-1}\Xi\bM M^{-1})_{ij},
\vs{12pt}\\

\commut{\pi_v^i}{\pi_v^j}&=&-i\hbar\dps{\f14}(M^{-1}G\Theta M^{-1})_{ij},&\hs{18pt}&\commut{u^i}{p^x_j}&=&i\hbar(M^{-1}\bM\bM M^{-1})_{ij},
\vs{12pt}\\

\commut{u^i}{u^j}&=&i\hbar(M^{-1}\Theta  M^{-1})_{ij},&\hs{18pt}&\commut{u^i}{\pi^v_j}&=&i\hbar\dps{\f12}(M^{-1}\Theta\bM M^{-1})_{ij},
\vs{12pt}\\

\commut{u^i}{p^u_j}&=&i\hbar\dps{\f12}(M^{-1}G    M^{-1})_{ij}&\hs{18pt}&\commut{p^x_i}{\pi_v^j}&=&i\hbar\dps{\f12}(M^{-1}G\bM M^{-1})_{ij},
\vs{12pt}\\

\commut{p^u_i}{p^u_j}&=&-i\hbar\dps{\f14}(M^{-1}\Xi GM^{-1})_{ij}&\hs{18pt}&\commut{p^x_i}{p^u_j}&=&-i\hbar\dps{\f12}(M^{-1}\bM\Xi\bM M^{-1})_{ij},
\vs{12pt}\\

\commut{x^i}{v_j}&=&i\hbar(M^{-1}\bM M^{-1})_{ij},&\hs{18pt}&\commut{\pi_v^i}{p^u_j}&=&i\hbar\dps{\f14}(M^{-1}G\bM M^{-1})_{ij},
\vs{12pt}\\

\commut{x^i}{\pi_v^j}&=&i\hbar\dps{\f12}(M^{-1}\bM\Theta M^{-1})_{ij}.&\hs{18pt}& & &

\end{array}
\eqno{(\mbox{D}1)}
$$

\newpage

\section{Different Type of Constraint Dynamical Model}

\ts{12pt}We shall here show that the noncommutative quantum system equivalent to $\mcS^*$, Eq.(3.24), can be constructed with  the model Lagrangian, which does not contain the redundant CCS $(u,p^u)$ and $(\lambda,\pi_{\lambda})$.

\subsection{Primary Hamiltonian System}

Let $L$ be the first-order singular Lagrangian defined with 
$$
\begin{array}{rcl}
L&=&L(x,\dot{x},v,\dot{v})\vs{6pt}\\
&=&\dps{\dot{x}^i\bM_{ij}v_j-\f12\dot{v}_i\Theta^{ij}v_j-\f12\dot{x}^i\Xi_{ij}x^j-h_0(x,v)},
\end{array}
\eqno{(\mbox{E}1)}
$$ 
where $h_0(x,v)$ also corresponds to the Hamiltonian in the final constraint quantum system $\mcS^*$.\\
\ts{12pt}Then, the initial unconstraint quantum system $\mcS=(\mcC,\mcA(\mcC),H(\mcC),\mcK)$ is given as follows: 
$$
\begin{array}{lcl}

\mcC&=&\{(x^i,p^x_i),(v_i,p_v^i)|i=1,\cdots,N\},\vs{6pt}\\

& &\mcA(\mcC):\commut{x^i}{p^x_j}=i\hbar\delta^i_j,\hs{12pt}\commut{v_i}{p_v^j}=i\hbar\delta_i^j,\hs{12pt}(\mbox{the others})=0,\vs{12pt}\\

H&=&\symmp{\mu^i_{(1)}}{\phi^{(1)}_i}+\symmp{\mu^i_{(2)}}{\phi^{(2)}_i}+h_{0}(x,v),\vs{12pt}\\

\mcK&=&\{\phi^{\mbtn{(1)}}_i,\phi^{\mbtn{(2)}}_i|i=1,\cdots,N\}\vs{6pt}\\
& &\phi^{\mbtn{(1)}}_i=\bM_{ij}v_j-p^x_i-\dps{\f12}\Xi_{ij}x^j,\hs{24pt}\phi^{\mbtn{(2)}}_i=\pi^i_v+\dps{\f12}\Theta^{ij}v_j,\vs{6pt}\\
\mcA(\mcK)&:&\commut{\phi^{\mbtn{(1)}}_i}{\phi^{\mbtn{(1)}}_j}=i\hbar\Xi_{ij},\vs{6pt}\\
& &\commut{\phi^{\mbtn{(1)}}_i}{\phi^{\mbtn{(2)}}_j}=i\hbar\bM_{ij},\vs{6pt}\\
& &\commut{\phi^{\mbtn{(2)}}_i}{\phi^{\mbtn{(2)}}_j}=i\hbar\Theta^{ij},\vs{12pt}\\

\mu_{\mbtn{(1)}}^i&=&-(M^{-1}\Theta M^{-1})_{ij}\partial^x_jh_0(x,v)-(M^{-1}\bM   
 M^{-1})_{ij}\partial_v^jh_0(x,v),\vs{6pt}\\

\mu_{\mbtn{(2)}}^i&=&-(M^{-1}\bM M^{-1})_{ij}\partial^x_jh_0(x,v)+(M^{-1}\Xi M^{-1})_{ij}\partial_v^jh_0(x,v),\vs{6pt}\\
& &(\mbox{Lagrange multiplier operators}).
\end{array}
\eqno{(\mbox{E}2)}
$$

\subsection{Projection of $\mcS$}

\subsubsection{ACCS $Z_{\alpha}$ and Projection Operator  $\Phat$ for $\mcK$}

Let $Z_{\alpha}(=\xi_i(\alpha=i),\e \pi_i(\alpha=i+N))$ be the ACCS corresponding to $\mcK$.
Following POM, then, the ACCS satisfies
$$
\begin{array}{l}
\xi_i-\dps{\f12}\Xi_{ij}\pi_j=\phi^{\mbtn{(1)}}_i,\vs{12pt}\\

\pi_i+\dps{\f12}\Theta^{ij}\xi_j=\phi^{\mbtn{(2)}}_i.
\end{array}
\eqno{(\mbox{E}3)}
$$
From Eqs.(\mbox{E}3), $\xi_i$ and $\pi_i$ are given as follows:
$$
\begin{array}{lcl}
\xi_i&=&M^{-1}_{ij}(\phi^{\mbtn{(1)}}_j+\dps{\f12}\Xi_{jk}\phi^{\mbtn{(2)}}_k)=M^{-1}_{ij}(v_j+\dps{\f12}\Xi_{jk}p_v^k-p^x_j-\dps{\f12}\Xi_{jk}x^k),\vs{12pt}\\

\pi_i&=&M^{-1}_{ij}(\phi^{\mbtn{(2)}}_j-\dps{\f12}\Theta_{jk}\phi^{\mbtn{(1)}}_k)=M^{-1}_{ij}(p_v^j+\dps{\f18}(G\Theta)_{jk}v_k+\dps{\f12}\Theta^{jk}p^x_k+\dps{\f14}G_{jk}x^k).

\end{array}
\eqno{(\mbox{E}4)}
$$
Let $\Phat$ be the projection operator constructed with $Z_{\alpha}$, which satisfies
$$
\Phat\Phat=\Phat,\hs{12pt}\Phat Z_{\alpha}=0,\hs{12pt}\Phat\Zp_{\alpha}=\Zm_{\alpha}\Phat=0.
\eqno{(\mbox{E}5)}
$$
 From Eqs.(\mbox{E}3), then, the projection conditions for $\mcK$ become as follows:
$$
\begin{array}{lcl}

\Phat\phi^{\mbtn{(1)}}_i(x,p^x,v)&=&\bM_{ij}\Phat v_j-\Phat p^x_i-\dps{\f12}\Xi_{ij}\Phat x^j=\phi^{\mbtn{(1)}}_i(\Phat x,\Phat p^x,\Phat v)=0,\vs{12pt}\\

\Phat\phi^{\mbtn{(2)}}_i(v,p_v)&=&\Phat p_v^i+\dps{\f12}\Theta^{ij}\Phat v_j=\phi^{\mbtn{(2)}}_i(\Phat p_v,\Phat v)=0.

\end{array}
\eqno{(\mbox{E}6)}
$$

\subsubsection{Projection of Operators} 

Let $\Phat\mcC$ be $\mcC^{\mbtn{(1)}}$,
$$
\mcC^{\mbtn{(1)}}=\Phat\mcC=\{(\Phat x,\Phat p^x),(\Phat v,\Phat p_v)\}=\{x,p^x,v,p_v\}.
\eqno{(\mbox{E}7))}
$$
Due to the projection conditions (E6), then, $\mcC^{\mbtn{(1)}}$ holds
$$
\begin{array}{l}

\bM_{ij}v_j-p^x_i-\dps{\f12}\Xi_{ij}x^j=0,\vs{6pt}\\

p_v^i+\dps{\f12}\Theta^{ij}v_j=0.

\end{array}
\eqno{(\mbox{E}8))}
$$
The commutator algebra of $\mcC^{\mbtn{(1)}}$, $\mcA(\mcC^{\mbtn{(1)}})$, becomes as follows:

$$
\begin{array}{rclcrcl}

\commut{x^i}{x^j}&=&i\hbar(M^{-1}\Theta  M^{-1})_{ij},&\hs{18pt}&\commut{x^i}{v_j}&=&i\hbar(M^{-1}\bM M^{-1})_{ij},
\vs{6pt}\\

\commut{x^i}{p^x_j}&=&i\hbar(M^{-1}(I+\dps{\f1{16}}G^2)M^{-1})_{ij},&\hs{18pt}&\commut{x^i}{p_v^j}&=&i\hbar\dps{\f12}(M^{-1}\bM\Theta M^{-1})_{ij},
\vs{6pt}\\

\commut{p^x_i}{p^x_j}&=&-i\hbar\dps{\f14}(M^{-1}G\Xi M^{-1})_{ij},&\hs{18pt}&\commut{v^i}{p^x_j}&=&i\hbar\dps{\f12}(M^{-1}\Xi\bM M^{-1})_{ij},
\vs{6pt}\\

\commut{v_i}{v_j}&=&i\hbar(M^{-1}\Xi M^{-1})_{ij},&\hs{18pt}&\commut{p^x_i}{p_v^j}&=&i\hbar\dps{\f14}(M^{-1}G\bM M^{-1})_{ij}.
\vs{6pt}\\

\commut{v_i}{p_v^j}&=&i\hbar\dps{\f12}(M^{-1}GM^{-1})_{ij},&\hs{18pt}& & &\vs{12pt}\\

\commut{p_v^i}{p_v^j}&=&-i\hbar\dps{\f14}(M^{-1}G\Theta M^{-1})_{ij},&\hs{18pt}& & &

\end{array}
\eqno{(\mbox{E}9)}
$$
\ts{12pt}From (E3), now, the projection of $\symmp{\mu^i_{(1)}}{\phi^{(1)}_i}$ is estimated as follows:
$$
\Phat\symmp{\mu^i_{(1)}}{\phi^{(1)}_i}=\Phat(\xipps{ }{i}\mu^i_{\mbtn{(1)}} -\dps{\f12}\Xi_{ij}\pipps{ }{j}\mu^i_{\mbtn{(1)}})=0,
$$
and similarly,
$$
\Phat\symmp{\mu^i_{(2)}}{\phi^{(2)}_i}=\Phat(\pipps{ }{i}\mu^i_{\mbtn{(2)}}+\dps{\f12}\Theta^{ij}\xipps{ }{j}\mu^i_{\mbtn{(2)}})=0.
$$   
Thus, the projection of $H$ is given by
$$
H^{\mbtn{(1)}}=\Phat H=\Phat h_{0}(x,v).
\eqno{(\mbox{E}10)}
$$

\subsection{Constraint Quantum Sysytem $\mcS^*$}

\ts{12pt}From the commutator algebra (E9), we shall adopt $\{(x,v)\}$ as the {\it projected} CCS $\mcC^*$ in $\mcS^*$:
$$
\mcC^*\{(x,v)\}=\{x^i,v_i|i=1,\cdots,N\},
\eqno{(\mbox{E}11)}
$$
with
$$
\begin{array}{l}

p^x_i=\bM_{ij}v_j-\dps{\f12}\Xi_{ij}x^j,\vs{12pt}\\

p_v^i=-\dps{\f12}\Theta^{ij}v_j.

\end{array}
\eqno{(\mbox{E}12)}
$$
Then, the canonical structure $\mcA^*(\mcC^*)$ is represented as
$$
\begin{array}{l}

\commut{x^i}{x^j}=i\hbar(M^{-1}\Theta  M^{-1})_{ij},\vs{6pt}\\

\commut{x^i}{v_j}=i\hbar(M^{-1}\bM M^{-1})_{ij},\vs{6pt}\\

\commut{v_i}{v_j}=i\hbar(M^{-1}\Xi M^{-1})_{ij},

\end{array}
\eqno{(\mbox{E}13)}
$$
and the projected Hamiltonian $H^*$ is given by
$$
H^*=\Phat h_0(x,v).
\eqno{(\mbox{E}14)}
$$
Thus, the constraint quantum system is constructed by
$$
\mcS^*=(\mcC^*,H^*).
\eqno{(\mbox{E}15)}
$$
\

\end{document}